\documentclass[
    ,final            
  ]
  {aipproc}

\layoutstyle{8x11single}

\begin{document}

\title{Off-Axis Afterglow Light Curves from High-Resolution Hydrodynamical
Jet Simulations}


\author{Hendrik J. van Eerten, Andrew I. MacFadyen and Weiqun Zhang}{
  address={Center for Cosmology and Particle Physics, Physics Department, New York University, New York, NY 10003}
}




\begin{abstract}
Numerical jet simulations serve a valuable role in calculating gamma-ray burst afterglow emission beyond analytical approximations. Here we present the results of high resolution 2D simulations of decelerating relativistic jets performed using the \textsc{ram} adaptive mesh refinement relativistic hydrodynamics code. We have applied a separate synchrotron radiation code to the simulation results in order to calculate light curves at frequencies varying from radio to X-ray for observers at various angles from the jet axis. We provide a confirmation from radio light curves from simulations rather than from a simplified jet model for earlier results in the literature finding that only a very small number of local Ibc supernovae can possibly harbor an orphan afterglow.

Also, recent studies have noted an unexpected lack of observed jet breaks in the \emph{Swift} sample. Using a jet simulation with physical parameters representative for an average \emph{Swift} sample burst, such as a jet half opening angle of 0.1 rad and a source redshift of $z = 2.23$, we have created synthetic light curves at 1.5 keV with artifical errors while accounting for \emph{Swift} instrument biases as well. A large set of these light curves have been generated and analyzed using a Monte Carlo approach. Single and broken power law fits are compared. We find that for increasing observer angle, the jet break quickly becomes hard to detect. This holds true even when the observer remains well within the jet opening angle. We find that the odds that a \emph{Swift} light curve from a randomly oriented 0.1 radians jet at $z = 2.23$ will exhibit a jet break at the $3\sigma$ level are only 12 percent. The observer angle therefore provides a natural explanation for the lack of perceived jet breaks in the \emph{Swift} sample.
\end{abstract}

\maketitle


\section{Introduction}

Numerical simulations of gamma-ray burst (GRB) afterglows allow us to probe in detail regions of parameter space that are unaccessible via simplified analytical models. For example, by integrating over the total synchrotron emission from a decelerating relativistic jet as calculated by a relativistic hydrodynamical (RHD) simulation or by solving the appropiate linear radiative transfer equations, it becomes possible to predict quantitatively the received flux for an observer positioned off the jet axis, without resorting to simplifying assumptions such as a homogeneous emission region or lateral spreading of the jet at fixed velocity.

We use a massively parallel adaptive-mesh refinement RHD code, \textsc{ram} \cite{Zhang2006} to simulate GRB afterglow jets. From the dynamical simulation results we calculate the received flux at various observer times and frequencies (from low radio to X-ray) using a separate synchrotron radiation module. We discuss this approach and general results for various jet parameters more extensively elsewhere in these proceedings\footnote{see \emph{An on-line library of afterglow light curves}, H.J. van Eerten, A.I. MacFadyen \& W. Zhang, elsewhere in these proceedings.}. The radiation code uses a parametrization smiliar to that in \cite{Sari1998} to describe the energy distribution of electrons accelerated at the shock front. For every grid cell in the data dumps from the fluid simulation this distribution is calculated and the resulting synchrotron  emission at a given observer frequency is stored at the appriate observer time bin. Alternatively, when synchrotron self-absorption plays a role, the radiative transfer equations for a dynamically varying number of rays are solved explicitly throughout the expanding fluid. These methods are explained in detail in \cite{Zhang2009, vanEerten2009, vanEerten2010, vanEerten2010c}.

In this contribution we discuss two applications of our numerical simulations plus radiation code. First we show that even a small observer angle (smaller than the jet half opening angle) already has a strong consequence for the observed flux. We demonstrate that for a typical \emph{Swift} \cite{Gehrels2004, Evans2007} afterglow, the jet break quickly becomes difficult to detect as we move off-axis. This explains the lack of perceived jet breaks noted in recent studies by various authors \cite{Racusin2009, Evans2009}. The analysis here largely follows that from \cite{vanEerten2010c}, albeit with an important difference: the light curves and conclusions drawn here are based on a similation with physical parameters that are typical for the \emph{Swift} sample (including a redshift $z = 2.23$ and jet half opening angle $\theta_h = 0.1$ rad ($5.7^{\circ}$), rather than those of \cite{vanEerten2010c}, where a wider jet was used ($\theta_h = 0.2$ rad) and the redshift was ignored.

Second, we show the synthetic radio light curves calculated for observers at very large angles, up to $90^{\circ}$. At these angles the prompt emission will not be registered by the observer and only an \emph{orphan afterglow} signal is expected. When comparing our detailed afterglow calculations to observationally established upper limits on local Ibc supernovae from \cite{Soderberg2006}, we confirm their conclusion that only a very small number of these supernovae can possibly harbor an orphan afterglow.

\section{Hidden \emph{Swift} jet breaks}

\begin{figure}
  \includegraphics[width=0.49\columnwidth]{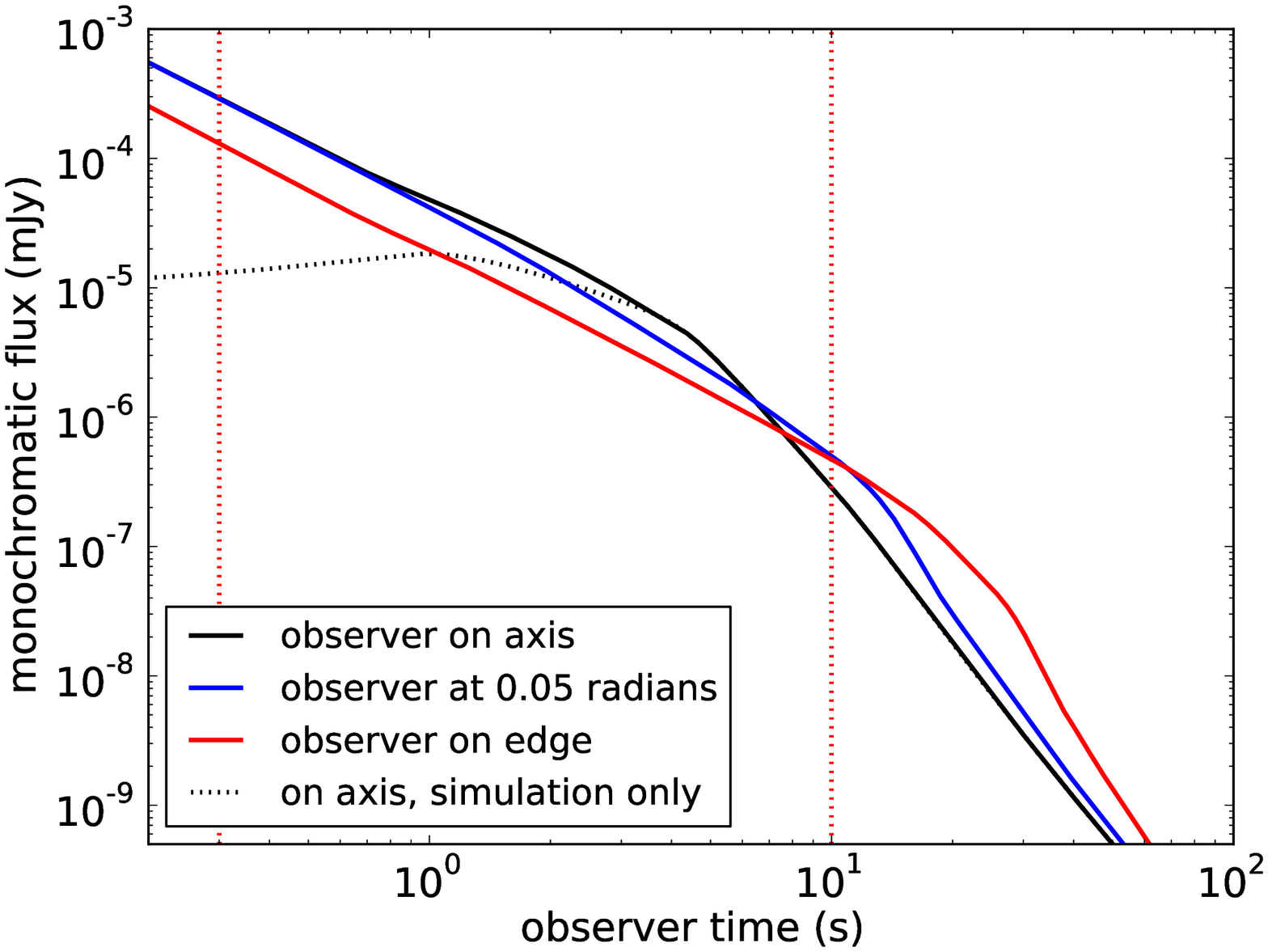}
  \includegraphics[width=0.49\columnwidth]{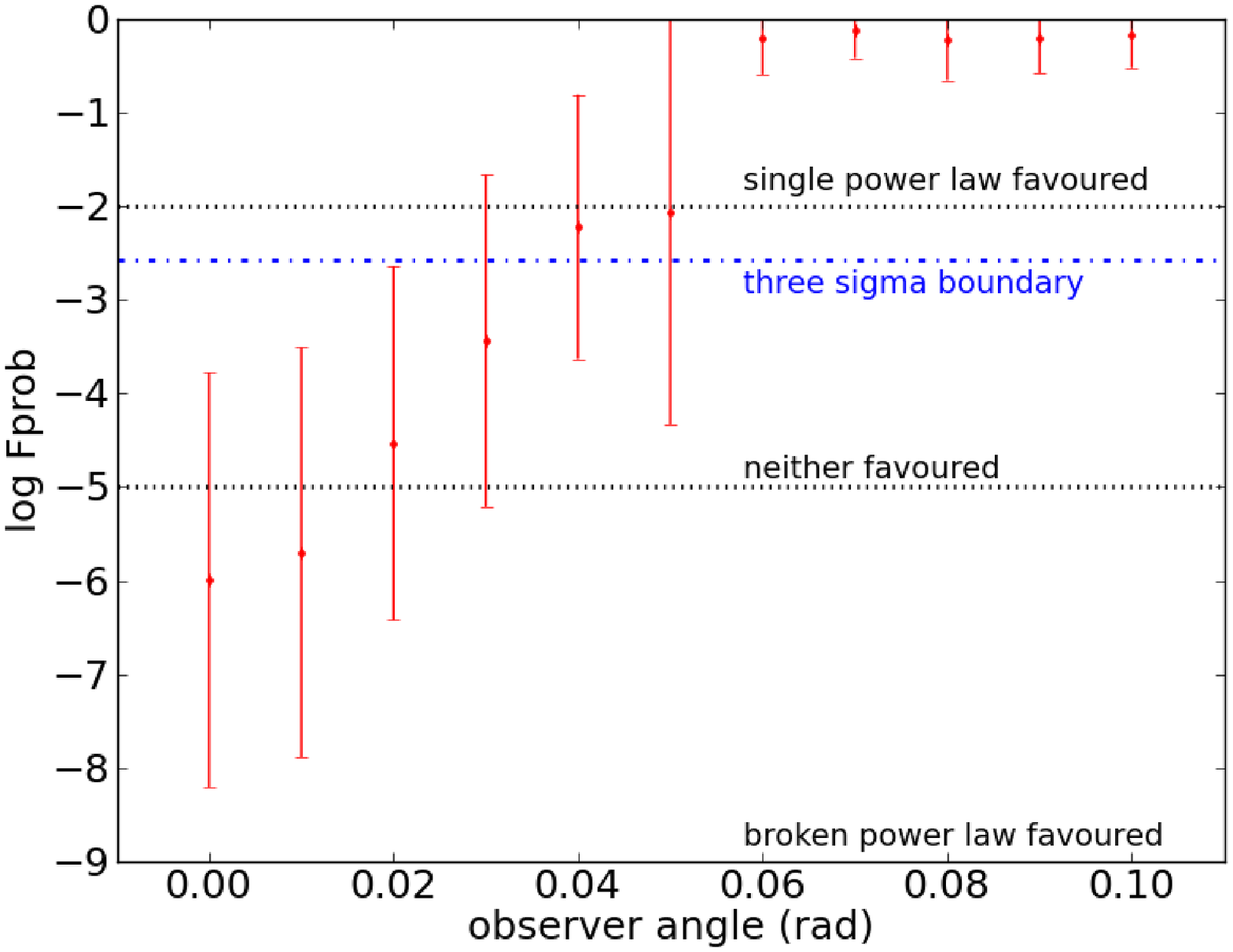}
  \caption{Left plot: Simulated X-ray light curves at 1.5 keV for observers on axis, halfway and on the jet edge, for a typical \emph{Swift} GRB with jet half opening angle 0.1 rad and redshift $z = 2.23$. Early time emission was added analytically assuming a the blastwave dynamics follow the Blandford-McKee solution \cite{Blandford1976}. \emph{Swift} data usually runs to $\sim 10 $ days, dictated by a lower sensitivity limit of $\sim 5 \times 10^{-4}$ cts s$^{-1}$. Right plot: The odds of finding a jet break by comparing single and double power law fits to synthetic \emph{Swift} light curves created from simulated light curves by adding artificial errors and instrument biases. The F test measures whether the improvement of a double power law fit over a single power law fit is significant. A large number of synthetic \emph{Swift} light curves is generated via a Monte Carlo approach and two examples are given in Fig. \ref{swift_curves_figure}.}
  \label{simulations_and_F_figure}
\end{figure}

\begin{figure}
  \includegraphics[width=0.49\columnwidth]{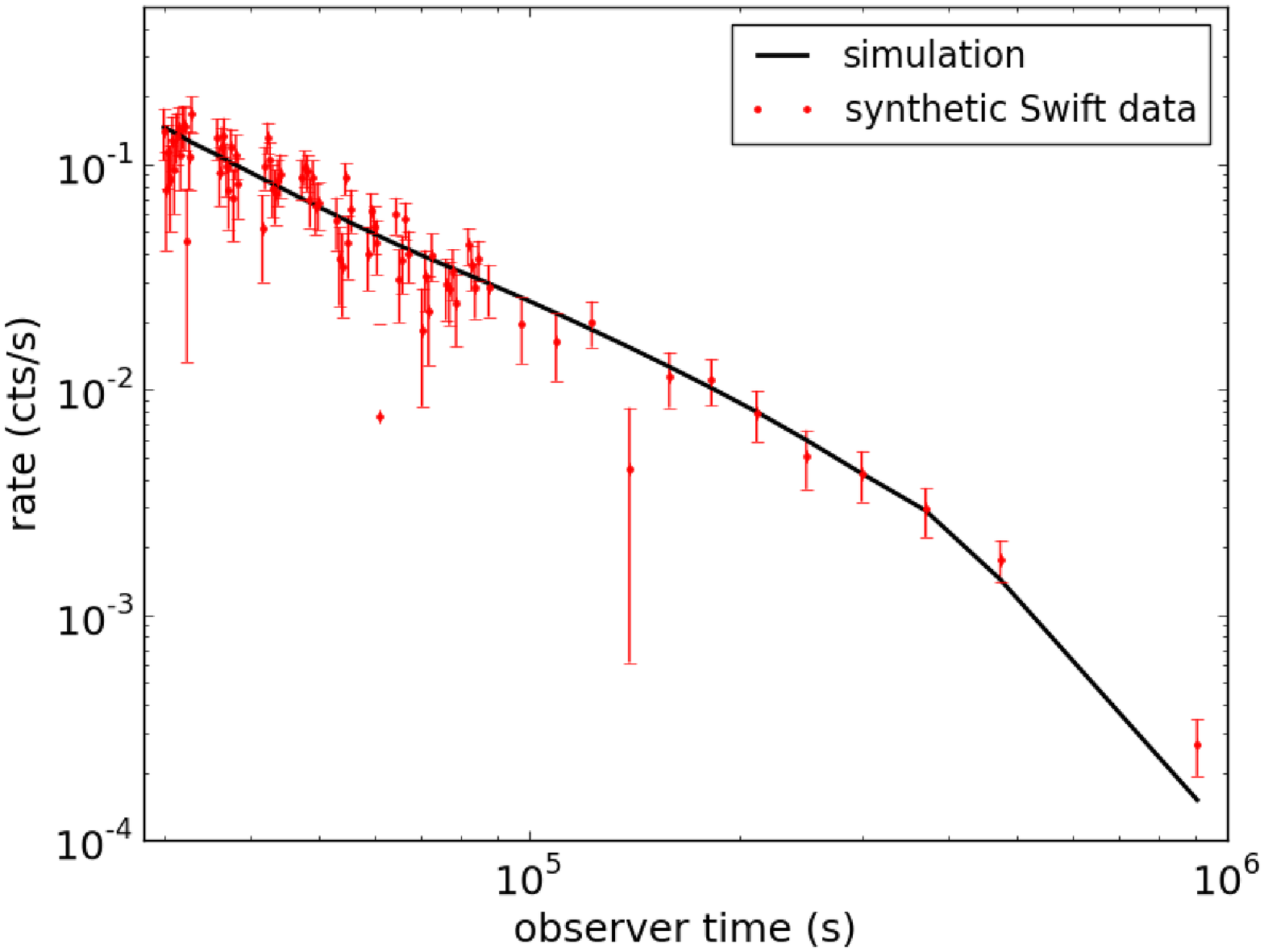}
  \includegraphics[width=0.49\columnwidth]{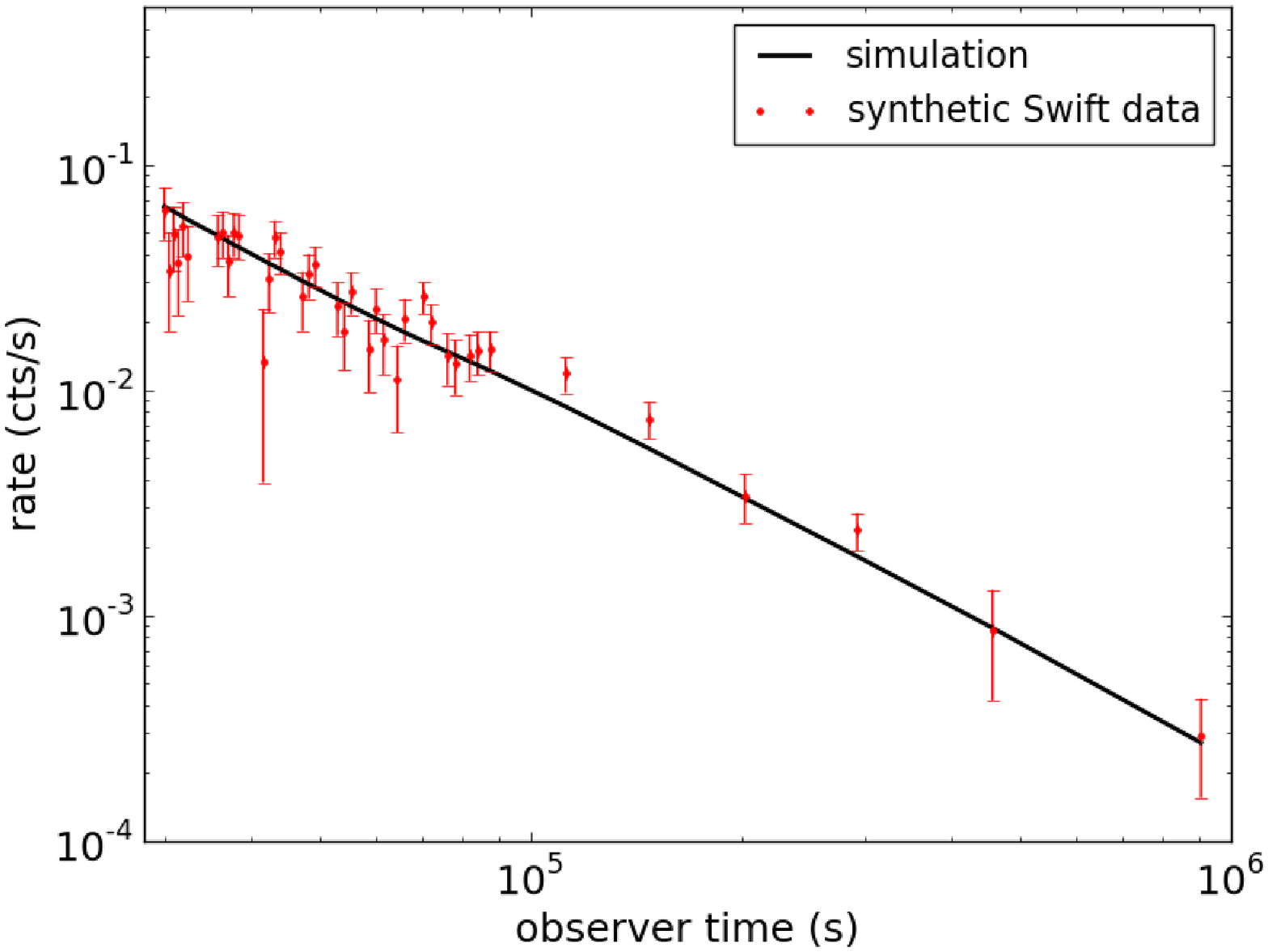}
  \caption{Two example synthetic \emph{Swift} light curves from the Monte Carlo generated sample of 1000 per observer angle. For each light curve the artificial errors are recalculated. Gaps due to \emph{Swift}'s low-Earth orbit and a decrease in fractional exposure from 1.0 to 0.1 after one day were added as instrument biases. }
  \label{swift_curves_figure}
\end{figure}

We have simulated afterglow jets with various physics parameters. Results for a half opening angle of 0.2 rad (11.5 degrees) have been published in \cite{vanEerten2010c}. Here we present, for the first time, results for afterglow parameters chosen to match those typically observed by Swift, with jet half opening angle of 0.1 rad (5.7 degrees), jet energy $2 \times 10^{51}$ ergs in both jets together and homogeneous circumburst number density of 1 cm$^{-3}$. The redshift is set to $z = 2.23$, and the observer angle is varied but kept within the jet opening angle. Light curves are calculated assuming that the magnetic energy density behind the afterglow jet shock front is equal to a fraction $\epsilon_B = 0.1$ of the local thermal energy density, the energy density of the accelerated particles is a fraction $\epsilon_E = 0.1$ of the local thermal energy density and the electrons have been accelerated to a power law distribution in energy with slope $p = -2.5$. Results for some key observer angles are shown in the left plot of Fig. \ref{simulations_and_F_figure}.

The resulting X-ray light curves form the starting point to generate synthetic Swift datasets. We translate monochromatic flux to counts per second and to a limited number of data points using realistic Swift settings (taken from \cite{Evans2007}). These settings are discussed in \cite{vanEerten2010c}. Our approach here differs however from the one presented there in one respect. Rather than rescaling the observed number of cts s$^{-1}$ to 0.1 at 1 day, combined with an unphysical 300 cts per bin in order to get light curves with an (observationally expected) number of $\sim 30$ data points, we take the more physical value of 30 cts per bin, apply an energy density for a single count of $3.8 \times 10^{11}$ erg cm$^{-2}$ and use the redshift $z = 2.23$ rather than the arbitrary rescaling at 1 day to set the number of cts s$^{-1}$. The average number of data points for a single curve is then again consistent with those in the \emph{Swift} sample. Example light curves are shown in Fig \ref{swift_curves_figure}.

A large number of datasets is generated with different random errors and to these datasets both single and broken power laws are fitted (a similar approach was taken in \cite{Curran2008} for analytically calculated curves seen on-axis only) These fit results are then compared using an F-test, to determine whether a broken power law fit yields a significant improvement over a single power law fit. This is shown in the right plot of Fig. \ref{simulations_and_F_figure}.

Our results explain why recent studies of the Swift sample \cite{Racusin2009, Evans2009} show a lack of afterglow jet breaks. For increasing observer angle, the jet break quickly becomes hard to detect -even when the observer remains well within the jet opening angle. The main cause for this is that the jet break gets postponed beyond what \emph{Swift} typically can observe (see Fig. \ref{simulations_and_F_figure}), although even before this happens the quality of the data minimizes the improvement of a double power law fit over a single power law fit. We emphasize again that these results have been obtained for a jet with physical parameters characteristic for the \emph{Swift} sample. For the current scenario, the odds that a \emph{Swift} light curve from a randomly oriented 0.1 radians jet at $z = 2.23$ will exhibit a jet break at the $3\sigma$ level are only 12 percent.

\section{Orphan afterglows}

\begin{figure}
  \includegraphics[width=0.49\columnwidth]{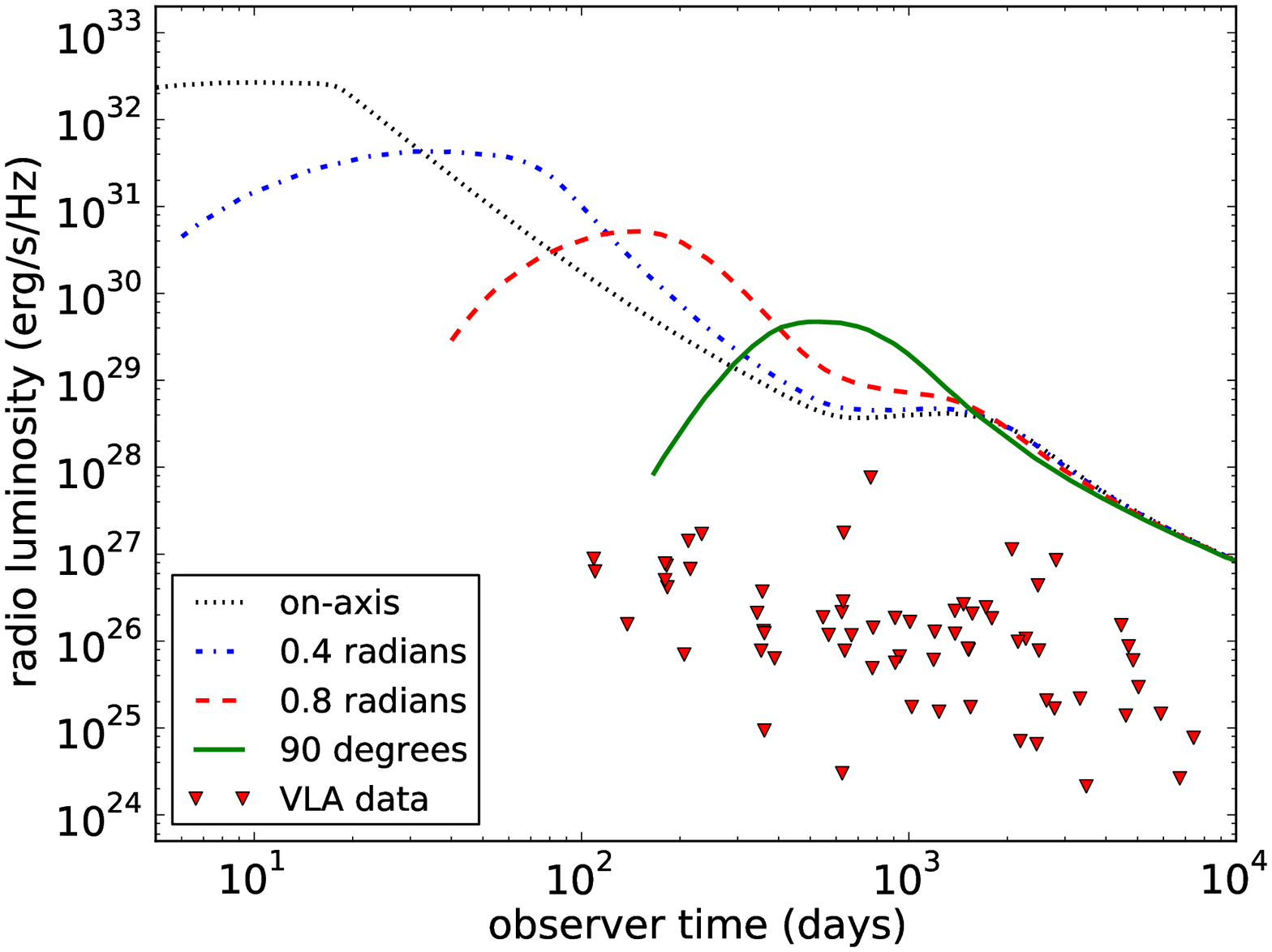}
  \includegraphics[width=0.49\columnwidth]{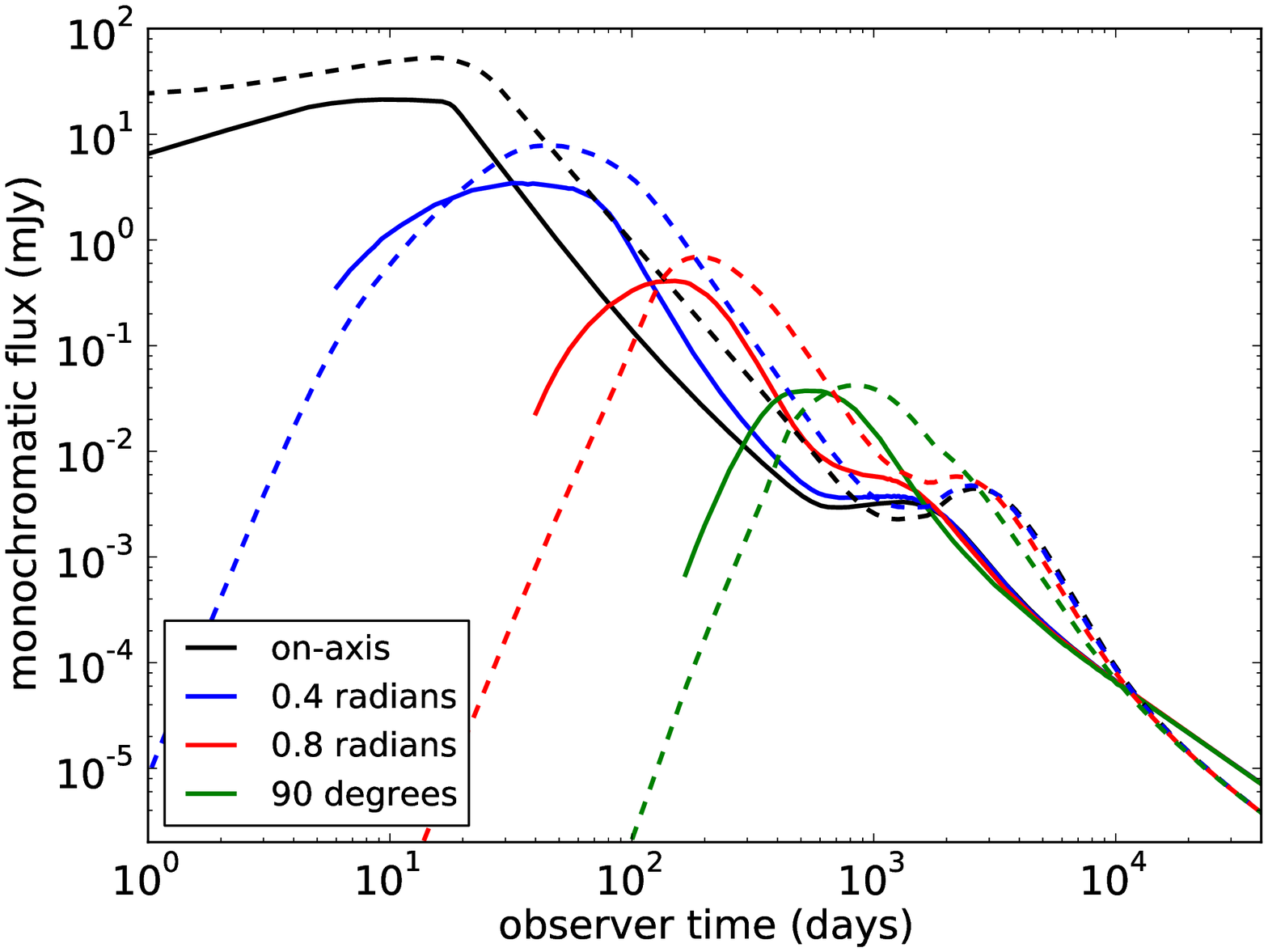}
  \caption{Left plot: VLA late time radio limits ($3 \sigma$) for 66 local type Ibc supernovae compared against simulation results for typical afterglow parameters. All supernova redshifts have been ignored (the largest redshift, that of SN 1991D, is $\sim 0.04$). The fluxes have been rescaled to luminosities.  All VLA observations were done at 8.46 GHz, the afterglow light curve is calculated at the same frequency. As in the rest of this paper, the simulation jet half opening angle is $11.5^\circ$ (0.2 rad). Right plot: Direct comparison between simulation results (solid lines) and an analytical model (dashed lines) for different observer angles, for radio frequency 8.46 Ghz. The analytical model assumes that the emission region is a homogeneous slab and that the jet starts spreading sideways at the speed of sound once it has reached a nonrelativistic velocity. Both the simulation and model light curves are calculated for the  jet half opening angle of 0.2 rad. The supernova data was taken from \cite{Soderberg2006}, the simulation data from \cite{vanEerten2010c}.}
  \label{orphan_afterglows_figure}
\end{figure}

The existence of orphan afterglows is an important and general prediction of current afterglow theories. Although various groups have looked for orphan afterglows, both in the optical and radio \cite{Levinson2002, Soderberg2006, Gal-Yam2006, Malacrino2007}, few positive detections have been reported and archival studies have mainly served to establish constraints on GRB rates and beaming factors. In \cite{Soderberg2006}, radio flux upper limits for a large set of Ibc type supernova observations are compared to an analytical estimate of the flux for a typical afterglow jet emitted from the same position and seen off-axis. They find that the expected signal from a typical afterglow is usually stronger than the radio upper limits by more than an order of magnitude, even for observers completely off-axis at $90^{\circ}$. The observed supernovae therefore show no signs of harboring an orphan afterglow. 

The study from \cite{Soderberg2006} uses a strongly simplified analytical description for the afterglow evolution. Our detailed numerical calculations confirm their result and show that the flux for observers far off-axis is even higher than expected from a simplified model based on the same physics. In the left plot of Fig. \ref{orphan_afterglows_figure} we show  a comparison between simulated afterglow light curves at $8.46$ Ghz and VLA late time radio limits for 66 local type Ibc supernovae. Both the light curves and supernovae are expressed as luminosities to allow for the comparison between the different supernovae and the light curves. The right plot shows a comparison between simulated light curves and light curves from a simplified analytical model analogous to the one used in \cite{Soderberg2006}. The main difference between our work and that of \cite{Soderberg2006} is that we use a wider jet half opening angle of 0.2 rad for the afterglow. We used a total energy in both jets of $2\cdot10^{51}$ erg, a homogeneous circumburst medium number density 1 g cm$^{-3}$, which is comparable to the settings from \cite{Soderberg2006}. Our simulation light curves for 0.2 half opening angle jets peak earlier than our model light curves for jets with the same opening angle and do so at lower peak luminosity. This is discussed in more detail in \cite{vanEerten2010c}, where this analysis was first presented.


\begin{theacknowledgments}This work was supported in part by NASA under Grant No. 09-ATP09-0190 issued through the Astrophysics Theory Program (ATP).  The software used in this work was in part developed by the DOE-supported ASCI/Alliance Center for Astrophysical Thermonuclear Flashes at the University of Chicago. We thank Peter Curran for allowing us the use of his computer code for synthesizing and fitting \emph{Swift} data.
\end{theacknowledgments}



\bibliographystyle{aipproc}   

\bibliography{offaxis}

\IfFileExists{\jobname.bbl}{}
 {\typeout{}
  \typeout{******************************************}
  \typeout{** Please run "bibtex \jobname" to optain}
  \typeout{** the bibliography and then re-run LaTeX}
  \typeout{** twice to fix the references!}
  \typeout{******************************************}
  \typeout{}
 }

\end{document}